\documentclass[a4paper]{jpconf}
\usepackage{graphicx}
\usepackage{textcomp}

\begin{document}
\title{Polycrystalline para-terphenyl scintillator adopted in 
a $\beta^-$ detecting probe for radio-guided surgery}

\author{E~Solfaroli~Camillocci$^{1,2}$,
F~Bellini$^{1,2}$,
V~Bocci$^{2}$,
F~Collamati$^{2,3}$, 
E~De~Lucia$^{4}$,
R~Faccini$^{1,2}$, 
M~Marafini$^{2,5}$,
I~Mattei$^{4,6}$,
S~Morganti$^{2}$, 
R~Paramatti$^{2}$, 
V~Patera$^{2,3}$,
D~Pinci$^{2}$,  
L~Recchia$^{2}$,
A~Russomando$^{1,2,7}$,
A~Sarti$^{3,4}$,
A~Sciubba$^{2,3}$,
M~Senzacqua$^{1,2}$ and C~Voena$^{2}$}

\address{
$^1$~Dip. Fisica, Sapienza Univ. di Roma, Rome, Italy;
$^2$~INFN Sezione di Roma, Rome, Italy;
$^3$~Dip. Scienze di Base e Applicate per l'Ingegneria, 
Sapienza Univ. di Roma, Rome, Italy;
$^4$~Laboratori Nazionali di Frascati dell'INFN, Frascati (RM), Italy;
$^5$~Museo Storico della Fisica e Centro Studi e Ricerche ``E. Fermi'', Rome, Italy;
$^6$~Dipartimento di Fisica, Universit\`a Roma Tre, Rome, Italy;
$^7$~Center for Life Nano Science@Sapienza, Istituto Italiano di Tecnologia, Rome, Italy.}
\ead{elena.solfaroli@roma1.infn.it}

\begin{abstract} 
A radio-guided surgery technique exploiting 
$\beta^-$ emitters is under development.
It aims at a higher target-to-background activity ratio
implying both a smaller radiopharmaceutical activity and
the possibility of extending the technique to cases 
with a large uptake of surrounding healthy organs. 
Such technique requires a dedicated
intraoperative probe detecting $\beta^-$ radiation.
A first prototype has been developed 
relying on the low density and high light yield of 
the diphenylbutadiene doped para-therphenyl organic scintillator.
The scintillation light 
produced in a cylindrical crystal, 5~mm in diameter and 3~mm in height,
is guided to a photo-multiplier tube by optical fibres. 
The custom readout electronics is designed to optimize its usage 
in terms of feedback to the surgeon, portability and 
remote monitoring of the signal.
Tests show that 
with a radiotracer activity comparable to those 
administered for diagnostic purposes
the developed probe can detect 
a 0.1~ml cancerous residual of meningioma in a few seconds.
\end{abstract}

\section{Introduction}

A novel radio-guided surgery (RGS) technique 
using $\beta^-$ radiation 
is being developed~\cite{Beta-RGS,PatentBeta-RGS}.
The main advantage with respect to 
the traditional RGS with $\gamma$ radiation~\cite{GammaRGS}
is the lower target-to-background activity ratio (TBR). 
$\beta^-$ radiation indeed penetrates only a few millimetres of tissue 
with essentially no $\gamma$ contamination
being the \textit{bremsstrahlung} contribution almost negligible.
Low background from healthy tissue close to the lesion allows 
a smaller radiotracer activity to detect tumour remnants, 
and the possibility of extending the technique to cases 
with a significant uptake from healthy organs around the lesion.

In the past,
the use of $\beta^+$ decaying tracers was proposed~\cite{Beta+Probe},
since they are largely diffused in diagnostic applications
(\textit{Positron Emission Tomography}, PET). 
The emitted positrons have a limited penetration and 
their detection is local, 
but at rest annihilate with electrons 
producing $\gamma$s with energy of 511~keV: 
the background persists and, actually, increases in energy
becoming more penetrating.
This approach has been studied in pre-clinical tests 
but it is not yet in use in clinical practice.
The largest limitations range from 
the activity to be administered to achieve a quick probe response, 
to the complexity of the probe 
and to the exposure of the medical personnel.

On the contrary,
pure $\beta^-$ emitting tracers
allow to develop a handy and compact probe which, 
being sensitive only to particles emitted locally 
and operating in a low background environment,
provides a clearer delineation of margins of the lesioned tissue.

Meningioma brain tumour was identified 
as the first clinical case of interest 
because of its sensitivity to DOTATOC~\cite{MeningiomaUptake},
a synthetic analogue of somatostatin,
that can be marked with the pure $\beta^-$ radionuclide
$^{90}$Y~\cite{90Y-DOTATOC}.
This isotope is suitable for $\beta^-$ RGS 
because of its half-life (64~h), 
electron energy spectrum (endpoint: 2.28~MeV) 
and absence of $\gamma$ emission.
A study~\cite{MeningiomaRGS} of the potentiality 
of the proposed RGS technique has been performed
on PET diagnostic exams of patients affected by meningioma 
after administration of $^{68}$Ga-DOTATOC
(the uptake for the tracer can be assumed independent 
from the linked radionuclide).
The study shows that a TBR
greater than 10 is observed in almost all cases,
making the effectiveness of the technique very promising.
To implement this RGS technique, 
we are developing an intraoperative probe detecting $\beta^-$ decays
based on para-terphenyl scintillator.
In this article we describe the first prototype
and its performance.

\section{The $\beta^-$ intraoperative probe}

The first prototype has been designed 
having in mind the constraint coming from the meningioma clinical case. 
The instrument is optimized to provide a very small device
with high sensitivity to low energy electrons, 
directional detection and a fast response (few seconds) 
compatible with the surgical activity.
The required target spatial resolution is fixed 
by the typical volume (0.1~ml) of a surgical tumour residual 
in this clinical case.

The radiation sensitive element of the $\beta^-$ probe 
is a small scintillator tip 
made of commercial poly-crystalline para-terphenyl 
doped to 0.1\% in mass with diphenylbutadiene~\cite{Budakovsky},
manufactured by Detec-Europe.
This material was adopted after a detailed study~\cite{PTerf} 
due to its high light yield 
($\sim$3 times larger than typical organic scintillators)
and non-hygroscopic property, 
while its low density minimizes the sensitivity to photons.
Different detector sizes have been tested
(see~\ref{sec:probeDesign}),
the best configuration is a cylinder of 5~mm in diameter
and 3~mm in height.
A black PVC ring with external diameter of 11~mm
encloses the detector providing a
shield against radiation coming from the sides.
A 10~$\mu$m-thick aluminium front-end sheet
covers the scintillator to ensure the light tightness.
This assembly is mounted on top of
an easy-to-handle aluminium cylindrical body 
(diameter 8~mm and length 14~cm).

In order to avoid risk 
of patient coming into contact with electrical devices,
the scintillation light is guided by four 50~cm long optical fibres
outside the probe to a Hamamatsu H10721-210
photo-multiplier tube (PMT).
This photo-sensor module has 
an integrated high voltage power supply circuit 
requiring an input voltage as low as 5~V, 
making this device compatible with the surgical environment.
The read-out electronics and the logic board are housed 
in a compact box that wireless connects to a remote monitor 
to display the counting rate.

\subsection{Optimization of the scintillator design}
\label{sec:probeDesign}

Detection of non-penetrating low energy $\beta$ particles
does not require thick scintillators and 
the high light yield of the para-terphenyl 
largely compensates its short light attenuation length
(about 5~mm~\cite{PTerf}).

\begin{figure}[htbp]
 \begin{minipage}[t]{0.48\textwidth}
 \begin{center}
 \includegraphics [width=0.95\textwidth]{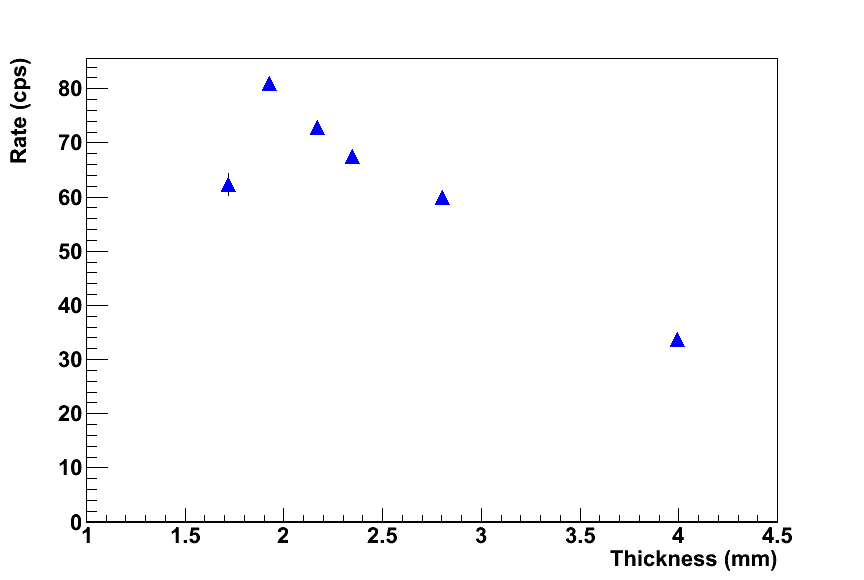}
 \caption{Probe counting rate as a function of the detector thickness.}
 \label{fig:Pter3mm}
 \end{center}
 \end{minipage}
  \hspace{0.03\textwidth}
\begin{minipage}[t]{0.48\textwidth}
 \begin{center}
\includegraphics[width=\textwidth]{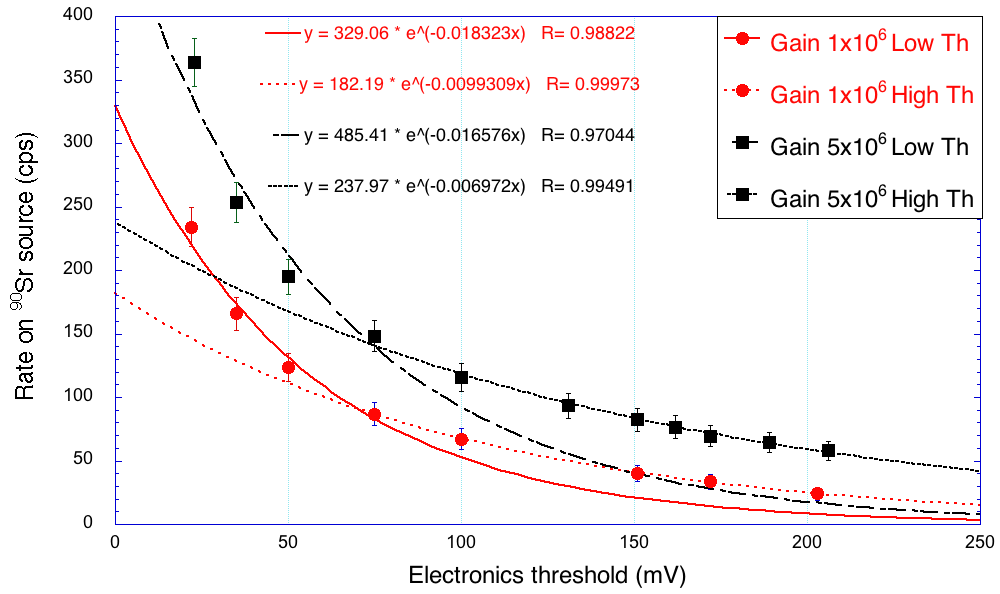}
 \caption{Probe counting rate vs electronics threshold. 
 Two PMT gain settings are shown. 
 In both cases after pulses are observed below 75~mV.}
 \label{fig:DarkCvsThr}
 \end{center}
\end{minipage}
\end{figure}

To maximize the sensitivity to $\beta$ particles,
the optimal thickness of the scintillator
has been determined experimentally.
A cylinder of para-terphenyl with diameter of 2.1~mm
is connected to the PMT by one optical fibre and
the thickness of this scintillator is gradually reduced
starting from 4~mm until the counting rate 
on a point-like $^{90}$Sr source with activity of 370~Bq
(\textit{Sr-source} in the following) 
is maximized.
The $^{90}$Sr 
is a long-lived $\beta^-$ emitter (half-life: 29 years)
with decay energy of 546~keV and no gamma radiation. 
It exists in secular equilibrium with its daughter isotope $^{90}$Y
resulting in the emission of two $\beta^-$ particles.
The radioactive source is sealed inside a 0.7~mm thick plastic disc.
In Fig.~\ref{fig:Pter3mm} the counting rate stored on the \textit{Sr-source}
as a function of the para-terphenyl thickness is shown.
The optimal point is found to be around 2~mm.

\subsection{The probe working point}

The probe working point is set by looking at the dark count 
and signal rates measured on the \textit{Sr-source} 
by fixing the PMT gain value and varying the electronics threshold 
that trigger the pulse counter. 
The results are shown in Fig.~\ref{fig:DarkCvsThr} 
for two PMT gain settings:
with a gain of 5$\cdot10^{6}$ and a threshold of 80~mV 
the contribution of after pulses and dark current 
is below 0.2~cps with a signal rate of 140~cps. 

\begin{figure}[htbp]
 \begin{center}
 \includegraphics [width=0.47\textwidth]{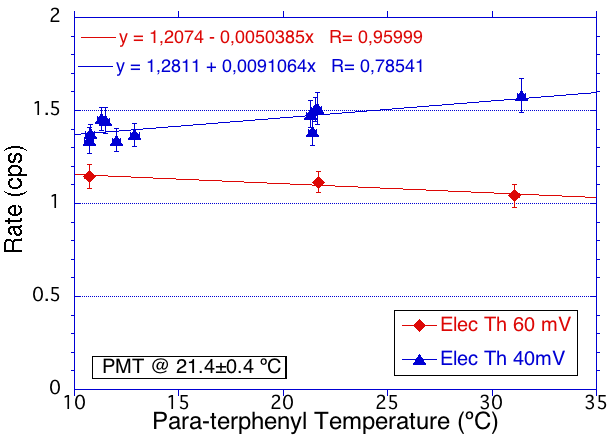}
 \hspace{0.03\textwidth}
 \includegraphics [width=0.47\textwidth]{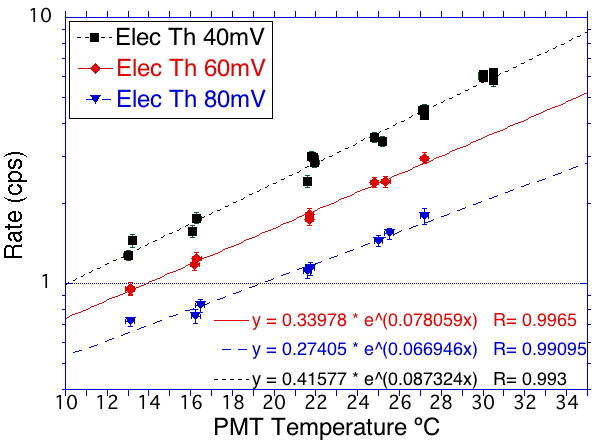}
 \caption{Test on stability of the probe electronic noise rate
 to temperature variations in the range 10-30~\textcelsius~for 
 the scintillator (left) and the PMT (right) separately.
 Configurations with different electronics thresholds are investigated.}
 \label{fig:DarkCvsT}
 \end{center}
 \end{figure}

The response stability to temperature variation 
in the range 10-30~\textcelsius~has been measured separately 
for the scintillating crystal and the PMT. 
As shown in Fig.~\ref{fig:DarkCvsT} 
the scintillator is almost insensitive to temperature variations
whereas the PMT dark counts increase 
according to the Hamamatsu specification.

\section{$\beta^-$ detection sensitivity}

The probe sensitivity in detecting electrons from $\beta^-$ decays
is measured with the \textit{Sr-source} (nominal activity 370~Bq).
The rate measured by the probe is 3.8$\cdot$10$^{5}$~cps/MBq
resulting in a detection efficiency of 40\%
of the $\beta^-$ decay products of the chain $^{90}$Sr/$^{90}$Y.

 \begin{figure}[htbp]
 \begin{minipage}[t]{0.49\textwidth}
 \begin{center}
 \includegraphics [width=\textwidth]{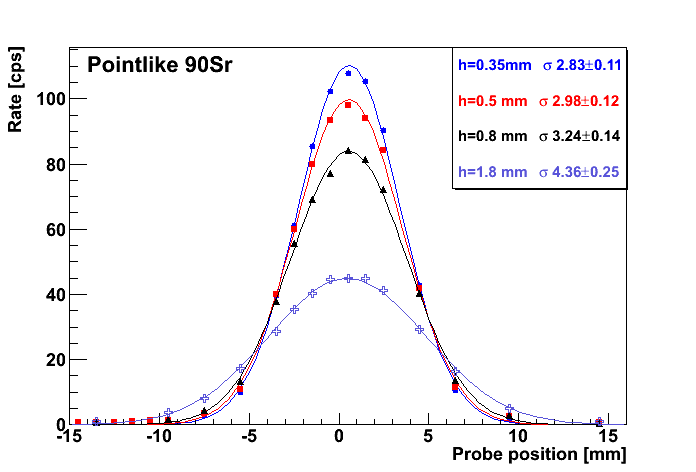}
 \caption{Profiles of the point-like \textit{Sr-source} 
 reconstructed by the probe at different distances
 from the source.}
 \label{fig:SrProfile}
 \end{center}
  \end{minipage}
    \hspace{0.02\textwidth}
  \begin{minipage}[t]{0.49\textwidth}
  \begin{center}
  \includegraphics [width=0.85\textwidth]{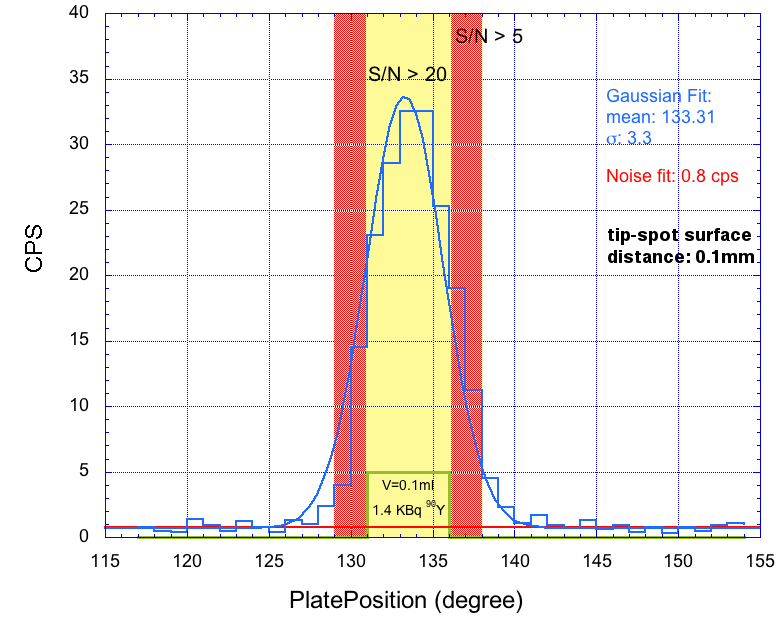}
  \caption{Lateral scan of a 1.4~kBq, 0.1~ml, $^{90}$Y phantom.  
  Gaussian and linear fits are
  for the signal and dark count rate respectively.
  See text for the band definition.}
  \label{fig:HotSpotProfile}
  \end{center}
  \end{minipage}
 \end{figure}
To study the capability of the probe
to identify active spots, 
the detector is mounted on a motorized linear actuator
ensuring position accuracy of 1.5~$\mu$m and
horizontal scans are performed over point-like and extended sources.
The scan over the point-like \textit{Sr-source} (Fig.~\ref{fig:SrProfile})
shows how spatial resolution depends on the distance 
between the probe tip and the surface.
The reconstructed profiles 
measured at distances ranging from 350~$\mu$m to 1.8~mm
are fitted using a Gaussian distribution
obtaining $\sigma$ between 2.8 and 4.4~mm.
The scan over the extended source (Fig.~\ref{fig:HotSpotProfile}),
 instead, explores the discovery power of the probe
in a more realistic simulation of a radio-labelled tumour residual.
A cylindrical phantom with volume of 0.1~ml
(diameter: 6.0~mm, height: 3.5~mm)
is filled with a saline solution with $^{90}$Y radionuclide
of 1.4~kBq activity,
comparable to those administered for diagnostic purposes.
The profile reconstructed by the probe is obtained
with an horizontal scan of 1.5~mm steps and 10~s per position and
a distance between the probe and the phantom surface of 100~$\mu$m.
The probe is able to well identify the active spot 
with a signal discriminating threshold S/N$>$5,
where S is the number of detected events and
N is the number of dark counts, and
the real phantom width is reconstructed with S/N$>$20.

\section{Background rejection}
 
\begin{figure}
\centering
\includegraphics[width=0.35\textwidth]{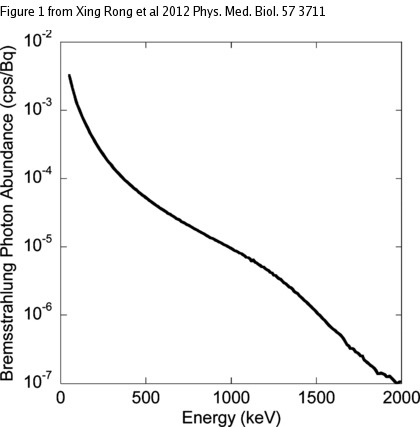}
\hspace{0.03\textwidth}
 \includegraphics [width=0.55\textwidth]{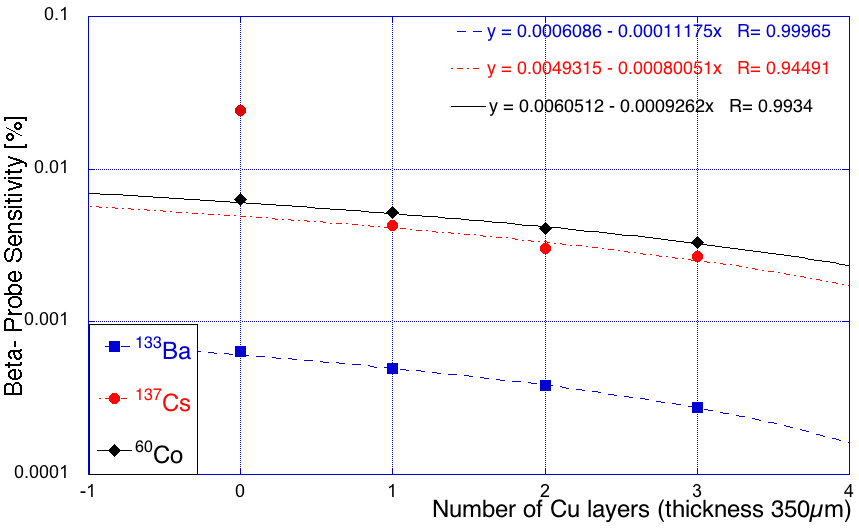}
\caption{\textbf{Left}: Energy spectrum (50-2000~keV, 10~keV interval) 
of $^{90}$Y \textit{Bremsstrahlung} photons in water.
\textbf{Right}: Probe sensitivity to photons
emitted by three different sources:
$^{60}$Co E$_{\gamma1}$=1170~keV E$_{\gamma2}$=1330~keV, 
$^{137}$Cs E$_\gamma$=662~keV,
$^{133}$Ba E$_\gamma$ ranging from 80 to 350~keV.
The measurements are repeated after insertions of
copper layers with thickness of 350~$\mu$m 
between the source and the probe tip 
to absorb the electronic component
of the Caesium emissions.}
\label{fig:Bremm90Y}
\end{figure}

The usage of pure $\beta^-$ emitting tracers
allows to operate with low radiation background, 
the residual being mainly due to photons 
coming from the \textit{Bremsstrahlung} radiation
of the electrons penetrating the tissue.
The abundance of the expected \textit{Bremsstrahlung} radiation 
as a function of the photon energy 
is shown in Fig.~\ref{fig:Bremm90Y} (left)
as computed with a simulation of
a $^{90}$Y source in water~\cite{90YBrem}.

We measure the sensitivity of the probe to these photons 
using three point-like sources:
$^{133}$Ba emitting photons with energy ranging from 80 to 350~keV,
$^{137}$Cs with gamma emission at E$_\gamma$=662~keV and
$^{60}$Co  with E$_{\gamma1}$=1170~keV and E$_{\gamma2}$=1330~keV.
To avoid signal from electrons in the case of Cs decays,
three copper layers with 350~$\mu$m thickness are inserted in sequence
between the source and the probe tip,
and the measurements are repeated at each step.
The counts as measured by the probe are shown 
in the Fig.~\ref{fig:Bremm90Y} (right) for the three sources.
Except for the first measure on the Cs source,
introducing the copper absorbers implies a very small decrease in rate 
compatible with the attenuation in copper and
the change in geometrical acceptance.

The sensitivity to the photons emitted by the $^{133}$Ba source
is below 10$^{-5}$, 
whereas for the $\gamma$s from the $^{137}$Cs and $^{60}$Co sources
the sensitivity is 
still below 10$^{-4}$.
These measurements allow to conclude that 
the intraoperative $\beta^-$ probe is not sensitive to 
the \textit{Bremsstrahlung} photons
and therefore the effectiveness of the RGS technique would
not be affected by this background.

\section{Conclusion}

A prototype of the intraoperative probe 
for RGS exploiting $\beta^-$ decays 
based on para-terphenyl scintillator 
has been developed and tested.
The $\beta^-$ detection efficiency
in the energy range of the $^{90}$Y emissions is $>$40\%, 
while the sensitivity to the \textit{Bremsstrahlung} photons is
below 10$^{-4}$
making this background radiation negligible. 
A test with a phantom simulating meningioma residual
shows that 
with a radiotracer activity comparable to those 
administered for diagnostic purposes
the probe is able to detect 
a 0.1~ml active spot in a few seconds.


\section*{References}
\begin{thebibliography}{9}

\bibitem{Beta-RGS}
Solfaroli Camillocci E \textit{et al}
2014
\textit{Sci. Rep.} \textbf{4} 4401 
DOI:10.1038/srep04401

\bibitem{PatentBeta-RGS}
Patent PCT/IT2014/000025 
deposited by Universit\`a degli studi di Roma ``La Sapienza'', 
Istituto Nazionale di Fisica Nucleare and 
Museo storico della fisica e centro studi e ricerche ``E. Fermi''.

\bibitem{GammaRGS}
Povoski SP \textit{et al}
2009
\textit{World Journal of Surgical Oncology} 7-11

\bibitem{Beta+Probe}
Bogalhas F \textit{et al}
2009
\textit{Phys Med Biol} \textbf{54} 4439-53

\bibitem{MeningiomaUptake}
Bartolomei M \textit{et al}
2009
\textit{Eur J Nucl Med Mol Imaging} \textbf{36} 1407-16

\bibitem{90Y-DOTATOC}
Heppeler A \textit{et al} 
1999
\textit{Chem Eur J} \textbf{7} 1974-81

\bibitem{MeningiomaRGS}
Collamati F \textit{et al}
2015
\textit{J Nucl Med} \textbf{56} 1-6 
DOI:10.2967/jnumed.114.145995 

\bibitem{glioblastomaUptake}
Heute D \textit{et al}
2010
\textit{J Nucl Med}  \textbf{51} 397-400

\bibitem{Budakovsky}
Budakovsky SV, Galunov NZ, Grinyov BV, Kim JK, Kim YK, Tarasenko OA
2009
\textit{Functional Materials} \textbf{16}, 1 86-91

\bibitem{PTerf}
Angelone M \textit{et al} 
2014
\textit{IEEE Transactions on Nuclear Science} \textbf{61}, 3 1483-87 
DOI:10.1109/TNS.2014.2322106

\bibitem{90YBrem}
Rong X, Du Y, Frey EC
2012
\textit{Phys Med Biol} \textbf{57} 3711–25 DOI:10.1088/0031-9155/57/12/3711

\end{thebibliography}
\end{document}